\documentclass[12pt]{article}
\usepackage{amsmath}
\usepackage{graphicx}
\usepackage{sectsty}
\newcommand{\bfOmega}{\mbox{\boldmath $\Omega$}}

\allsectionsfont{\centering}
\begin{document}

\title{Satellite Orbits and Relative Motion in Levi-Civita Coordinates}
                   
\author{Mayer Humi \\
Department of Mathematical Sciences\\
Worcester Polytechnic Institute\\
100 Institute Road\\
Worcester, MA  01609}

\maketitle
\thispagestyle{empty}

\begin{abstract}
In this paper we consider satellite trajectories in central force field with
quadratic drag using two formalisms. The first using polar coordinates
in which the angular momentum plays a dominant role. The second is in 
Levi-Civita coordinates in which the energy plays a central role. We then
unify these two formalisms by introducing polar coordinates in Levi-Civita
space and derive a new equation for satellite orbits in which energy and 
and angular momentum are on equal footing {and thus characterize the orbit 
by its two invariants}.
In the second part of the paper we derive in Levi-Civita coordinates
a linearized equation for the relative motion of two satellites whose 
trajectories are in the same plane.
We carry out also a numerical verification of these equations. 
\end{abstract}

\newpage
\setcounter{equation}{0}
\section{Introduction}

The accurate computation of satellites orbits has been the subject of 
numerous monographs [1-6] and research papers [7-9] (to name a few).
In particular the effect of the Earth oblateness [6,17,18]
and drag forces on satellite orbits [10,11,25] were the subject of some 
recent publications [19-21].

Since satellite orbits in a central force field are in a plane it is 
natural to use polar representation for the orbit equations. In this formalism
the angular momentum of the satellite emerges as a key variable for the
derivation of the equations of motion. However in 1920 [6,23] 
Levi-Civita introduced another two dimensional formalism whose primary 
objective was to regularize the equations of motion near collision 
which is advantageous from computational point of view.
To this end a coordinate transformation was defined and in the resulting 
equations of motion, energy emerges as the major quantity. A generalization
of Levi-Civita coordinates to three dimensions was made by 
Kustaanheimo and Steifel (KS-coordinates) [6,24,25] 

Related to the problem of satellite trajectory determination 
is the issue of relative motion of two satellites in orbit and the 
rendezvous problem.[12-16]. This problem was considered in several settings.
In particular the rendezvous problem in the non-central force field of 
an oblate body was addressed in [18]. However as far as we know 
this problem was not addressed even in two dimensions using Levi-Civita 
coordinates which mitigate to some extent the singular nature of these 
equations.

Our primary objective in this paper is to introduce polar coordinates in 
Levi-Civita plane and derive a new equation which charaterize the motion 
of a particle in a central force field in terms of its two invariants 
viz. energy and angular momentum. These invariants appear on equal footings 
in this equation. Thus this equation might find applications in classical
mechanics and in satellite theory, mission planing and control.
Our secondry objective in this paper, is to recast in Levi-Civita coordinates 
the relative motion of two satellites orbiting in the same plane thereby 
mitigating the sigular nature of these equation when these satelittes are 
in close proximity.

Within this framework we consider also the dissipative effects of quadratic 
drag on the the angular momentum and energy of a satellite and their
impact on it orbits. This problem was addressed in various context
by several authors [16,18,19,26,27] in the past.

The plan of the paper is as follows: In Sec $2$ we review the general 
theory for satellite trajectories in a central force field and derive 
the orbit equation under the action of quadratic drag.
Sec $3$ provides a short review of Levi-Civita formalism. In Sec $4$  
we introduce polar coordinates in Levi-Civita plane and derive the orbit 
equation in these coordinates. In Sec $5$ we derive a linearized equation 
for the relative motion of two satellites in the same plane. Sec $6$
carries out some numerical simulations to verify the accuracy of the
formulas that were derived in Secs $4,\,5$. We end up with some conclusions 
in Sec $7$. 

\setcounter{equation}{0}
\section{Angular Momentum Representation of the Orbit}
In this paper we consider satellites in a central force field with quadratic
drag. The general equation for the orbit of a satellite under these 
assumptions is 
\begin{equation}
\label{2.1}
{\ddot {\bf R}}=-f(R){\bf R}-g(\alpha,R)({\dot {\bf R}}\cdot 
{\dot {\bf R}})^{1/2} {\dot {\bf R}},\,\,\, R=|{\bf R}|
\end{equation}
In this equation {\bf R} is the radius vector of the satellite from the 
center of attraction and $\alpha$ is a parameter that lumps together drag
constant and the atmospheric density proportionality constants. 
Differentiation with respect to time is denoted by a dot. We assume that 
$f$ and $g$ are differentiable on the domain of $R$ under consideration.

Taking the vector product of (\ref{2.1}) with ${\bf R}$ on the left
and introducing the angular momentum vector
\begin{equation}
\label{2.2}
{\bf J}={\bf R} \times {\dot {\bf R}}
\end{equation}
we obtain
\begin{equation}
\label{2.3}
{\dot{\bf J}}=-g(\alpha,R)({\dot {\bf R}}\cdot {\dot {\bf R}})^{1/2}  {\bf J}
\end{equation}
It follows from this equation that ${\dot{\bf J}}$ is always parallel to 
${\bf J}$ and therefore ${\bf J}$ has a fixed direction. As a consequence
the motion is in a fixed plane which we take, without loss of generality, 
to be the $x-y$ plane. Introducing polar coordinates $R$ and $\theta$
in this plane the angular momentum vector can be written as
\begin{equation}
\label{2.4a}
{\bf J}=R^2{\dot\theta}{\bf e}_J
\end{equation}
where ${\bf e}_J$ is a unit vector in the direction of ${\bf J}$.
Hence (\ref{2.3}) can be rewritten as
\begin{equation}
\label{2.5}
\frac{{\dot J}}{J}=-g(\alpha,R)({\dot {\bf R}}\cdot {\dot {\bf R}})^{1/2}
\end{equation}
where $J=|{\bf J}|$.
Substituting this result in (\ref{2.1}) and dividing by $J$ yields
\begin{equation}
\label{2.6}
\frac{d}{dt}\left(\frac{{\dot{\bf R}}}{J}\right)+\frac{f(R){\bf R}}{J}=0
\end{equation}
i.e
\begin{equation}
\label{2.7}
{\dot{\bf R}}=-J\int\frac{f(R){\bf R}}{J}\,dt
\end{equation}

In polar coordinates (\ref{2.1}) becomes
\begin{equation}
\label{2.8}
R{\ddot\theta}+2{\dot R}{\dot \theta}=-g(\alpha,R)({\dot R}^2 +
R^2{\dot \theta}^2)^{1/2} R{\dot \theta}
\end{equation}
\begin{equation}
\label{2.9}
{\ddot R}-R{\dot \theta}^2=-f(R)R-
g(\alpha,R)({\dot R}^2 +R^2{\dot \theta}^2)^{1/2} {\dot R}
\end{equation}
Dividing the first equation by $R{\dot \theta}$ and integrating we obtain
\begin{equation}
\label{2.10}
J=R^2{\dot \theta}=h\exp{\left(-\int g(\alpha,R)({\dot R}^2 +
R^2{\dot \theta}^2)^{1/2} \, dt\right)}
\end{equation}
where $h$ is an integration constant.

Using (\ref{2.4a}) to change the independent variable from $t$ to $\theta$ 
we obtain after some algebra the orbit equation
\begin{equation}
\label{2.11}
\frac{R''}{R}-2\left(\frac{R'}{R}\right)^2+\frac{f(R)R^4}{J^2(\theta)}=1
\end{equation}
where primes denote differentiation with respect to $\theta$

Equation (\ref{2.11}) is  the orbit equation for the motion of the 
satellite in the "angular momentum representation".

When $f(R){\bf R} =\nabla V$ (\ref{2.1}) becomes
\begin{equation}
\label{2.12}
{\ddot {\bf R}}=-\nabla V -g(\alpha,R)({\dot {\bf R}}\cdot 
{\dot {\bf R}})^{1/2} {\dot {\bf R}}
\end{equation}
Taking the scalar product of this equation by ${\dot{\bf R}}$ we obtain
\begin{equation}
\label{2.13}
({\ddot {\bf R}},{\dot{\bf R}})=-(\nabla V,{\dot{\bf R}})-
g(\alpha,R)({\dot {\bf R}}\cdot{\dot {\bf R}})^{3/2}.
\end{equation}
This can be rewritten as
\begin{equation}
\label{2.14}
\frac{dE}{dt}=-g(\alpha,R)({\dot {\bf R}}\cdot{\dot {\bf R}})^{3/2}.
\end{equation}
where $E$ is the particle energy
$$
E=V+\frac{({\dot{\bf R}},{\dot{\bf R}})}{2}
$$
Eq. (\ref{2.14}) gives the rate of the particle energy decay due to the
dissipative effects of the drag force.

\setcounter{equation}{0}
\section{Energy Representation of the Orbit}

Energy is an invariant which characterize satellite motion
when no drag is present. To take advantage of this fact Levi-Civita[ ]
introduced a two dimensional formalism (which was latter generalized to 
three dimensions[]) in which Energy plays a central role (and has additional
advantages when collisions are present).

We present here a short summary of this formalism [ ]
\subsection{Levi-Civita formalism}

There are in the literature excellent expositions of Levi-Civita formalism [6]. 
Here we present a short overview of this formalism for completeness.

To begin with introduce a "fictitious time" $s$ which is
defined by the relation
\begin{equation}
\label{3.1}
\frac{d}{ds}=R\frac{d}{dt}.
\end{equation}
It is then easy to see that for ${\bf R}=(x,y)$
\begin{equation}
\label{3.2}
{\ddot{\bf R}}=\frac{1}{R^2}{\bf R}^{\prime\prime}-
\frac{1}{R^3}R^{\prime}{\bf R}^{\prime}.
\end{equation}
where primes denote differentiation with respect to $s$.
Furthermore the velocity $v$ satisfies,
\begin{equation}
\label{3.3}
v^2=({\dot{\bf R}},{\dot{\bf R}})=
\frac{1}{R^2}({\bf R}^{\prime},{\bf R}^{\prime})
\end{equation}
Using (\ref{3.2}) and (\ref{3.3}) the equation of motion (\ref{2.1}) becomes
\begin{equation}
\label{3.4}
\frac{1}{R^2}{\bf R}^{\prime\prime}-
\frac{1}{R^3}R^{\prime}{\bf R}^{\prime}=-f(R){\bf R} -\frac{g(\alpha,R)}{R^2}
\left({\bf R}^{\prime},{\bf R}^{\prime}\right)^{1/2}{\bf R}^{\prime}
\end{equation}

For a particle of unit mass whose equation of motion is (\ref{2.1}) where
$f(R){\bf R}=\nabla V$ and $g(\alpha,R)=0$ (no drag) the energy
$E$ is conserved and
\begin{equation}
\label{3.5}
E=V+\frac{v^2}{2}=V+\frac{1}{2R^2}({\bf R}^{\prime},{\bf R}^{\prime}).
\end{equation}
When this particle is in the gravitational field of a point mass $M$,
\begin{equation}
\label{3.6}
f(R)=\frac{\mu}{R^3},\,\,\, V=-\frac{\mu}{R}
\end{equation}
and
\begin{equation}
\label{3.7}
E=-\frac{\mu}{R}+\frac{1}{2R^2}({\bf R}^{\prime},{\bf R}^{\prime}),
\end{equation}
where $\mu=GM$ and $G$ is the gravitational constant. The rate of change in 
$E$ in this coordinate system can be obtained from (\ref{2.14}) by a change 
of variables
\begin{equation}
\label{3.7a}
\frac{dE}{ds}=-\frac{g(\alpha,R)}{R^{2}}({\bf R}',{\bf R}')^{3/2}
\end{equation}
Next define a transformation from the $(x,y)$ coordinate system to a new 
one $(u_1,u_2)$ (Levi-Civita coordinates) which is defined by the 
following relations,
\begin{equation}
\label{3.8}
x=u_1^2-u_2^2,\,\,\, y=2u_1u_2.
\end{equation}
(We shall refer to this as the U-plane).

Introducing the scalar product of any two vectors ${\bf w_1}$, ${\bf w_2}$ as 
$$
({\bf w_1},{\bf w_2})={\bf w_1}^T{\bf w_2},
$$
it follows that $R=({\bf u},{\bf u})=|{\bf u}|^2$ where ${\bf u}=(u_1,u_2)$.
Next we introduce Levi-Civita matrix,
\begin{eqnarray}
\label{3.9}
L({\bf u})=\left(\begin{array}{cc}
u_1 & -u_2 \\ \notag
u_2 &  u_1 
\end{array}\right).
\end{eqnarray}
Observe that the transpose $L^T$ and the inverse $L^{-1}$ of this matrix 
satisfy the following relationships
\begin{equation}
\label{3.10}
L^T({\bf u})L({\bf u})=R{\bf I},\,\,\, L^{-1}({\bf u})=\frac{1}{R}L^T({\bf u})
\end{equation}
where ${\bf I}$ is the unit two dimensional matrix.
Moreover for any two vectors ${\bf u},{\bf v}$ we have the following identities
\begin{equation}
\label{3.11}
L({\bf u}){\bf v}=L({\bf v}){\bf u},\,\,\,
({\bf u},{\bf u})L({\bf v}){\bf v}+({\bf v},{\bf v})L({\bf u}){\bf u}=
2({\bf u},{\bf v})L({\bf u}){\bf v}
\end{equation}
and
\begin{equation}
\label{3.12}
L({\bf u})^{\prime}=L({\bf u^{\prime}})
\end{equation}
It is then easy to see that for ${\bf R}$
\begin{equation}
\label{3.13}
{\bf R}=L({\bf u}){\bf u},\,\,\, {\bf R}^{\prime}=2L({\bf u}){\bf u}^{\prime},
\,\,\,{\bf R}^{\prime\prime}=2L({\bf u}){\bf u}^{\prime\prime}+
2L({\bf u})^{\prime}{\bf u}^{\prime}=2L({\bf u}){\bf u}^{\prime\prime}+
2L({\bf u}^{\prime}){\bf u}^{\prime}
\end{equation}
To convert (\ref{3.4}) to an equation in U space we use (\ref{3.13}) and
(\ref{3.11}) with the vectors ${\bf u},{\bf u}^{\prime}$. After some algebra
we obtain
\begin{equation}
\label{3.14}
{\bf u}^{\prime\prime}-\frac{({\bf u}^{\prime},{\bf u}^{\prime})}
{({\bf u},{\bf u})}{\bf u}=\frac{({\bf u},{\bf u})^2}{2}
\left\{-f(R){\bf u}-4g(\alpha,R)({\bf u},{\bf u})^{-3/2}
({\bf u}^{\prime},{\bf u}^{\prime})^{1/2} {\bf u}^{\prime}\right\}
\end{equation}
where $R$ has to be replaced by $({\bf u},{\bf u})$.
When $f(R)$ is given by (\ref{3.6}) this equation becomes
\begin{equation}
\label{3.15}
{\bf u}^{\prime\prime}+\frac{1}{2}\frac{\mu-
2({\bf u}^{\prime},{\bf u}^{\prime})}
{({\bf u},{\bf u})}{\bf u}=-2g(\alpha,R)({\bf u},{\bf u})^{1/2}
({\bf u}^{\prime},{\bf u}^{\prime})^{1/2} {\bf u}^{\prime}
\end{equation}
When no drag forces are present the particle energy eq. (\ref{3.7}) becomes
\begin{equation}
\label{3.16}
E=-\frac{\mu-2({\bf u}^{\prime},{\bf u}^{\prime})}{({\bf u},{\bf u})}.
\end{equation}
Using this expression for $E$ in (\ref{3.15}) we have
\begin{equation}
\label{3.17}
{\bf u}^{\prime\prime}-\frac{1}{2}E{\bf u}=
-2g(\alpha,R)({\bf u},{\bf u})^{1/2}
({\bf u}^{\prime},{\bf u}^{\prime})^{1/2} {\bf u}^{\prime}.
\end{equation}
However observe that due to the dissipative nature of the drag force,
$E$ is not constant under the present settings.

\setcounter{equation}{0}
\section{Polar Representation of the Orbit Equation}

Polar coordinates representation of satellite orbit equations in the $x-y$
plane offers several advantages over their Cartesian counterparts. Motivated by 
this observation we develop in this section a representation of (\ref{3.15}) 
using polar coordinates using Levi-Civita coordinates.

\subsection{Polar geometry in the U-plane}

We introduce polar coordinates $(u,\phi)$ in the U-plane in a manner 
similar to polar  coordinates in the the $x-y$ plane.
\begin{equation}
\label{4.1}
u=({\bf u},{\bf u})^{1/2},\,\,\, \phi=tan^{-1} \frac{{u_2}}{{u_1}}.
\end{equation}
The relationship between these variables and $(R,\theta)$ is given by
\begin{equation}
\label{4.2}
R=({\bf u},{\bf u})=u^2,\,\,\, \theta=2\phi.
\end{equation}
Furthermore in parallel to the definitions of the radial and tangential 
unit vectors in $x-y$ plane
\begin{equation}
\label{4.3}
{\bf e}_r=(\cos\theta,\sin\theta),\,\,\, {\bf e}_{\theta}=
(-\sin\theta,\cos\theta),
\end{equation}
we define in the U-plane
\begin{equation}
\label{4.4}
{\bf e}_u=(\cos\phi,\sin\phi),\,\,\, {\bf e}_{\phi}=
(-\sin\phi,\cos\phi).
\end{equation}
Using (\ref{4.2}) we find that
\begin{equation}
\label{4.5}
{\bf e}_r=\cos\phi\,{\bf e}_u+\sin\phi\,{\bf e}_{\phi},\,\,\, 
{\bf e}_{\theta}=-\sin\phi\,{\bf e}_u+\cos\phi\,{\bf e}_{\phi}.
\end{equation}

Since ${\bf u}=u{\bf e}_u$ we have the following formulas for the 
derivatives of ${\bf u}$,
\begin{equation}
\label{4.6}
{\bf u}^{\prime}=u^{\prime}{\bf e}_u+u\phi^{\prime}{\bf e}_{\phi},\,\,\,\,
{\bf u}^{\prime\prime}=(u^{\prime\prime}-u(\phi^{\prime})^2){\bf e}_u+
(u\phi^{\prime\prime}+2u^{\prime}\phi^{\prime}){\bf e}_{\phi}.
\end{equation}

\subsection{Derivation of the New Orbit Equation}

Using (\ref{4.6}) the orbit equation (\ref{3.14}) becomes
\begin{eqnarray}
\label{4.7}
&&[u''-u(\phi')^2]{\bf e}_u+[u\phi''+2u'\phi']{\bf e}_{\phi}-
\frac{({\bf u}',{\bf u}')}{u}{\bf e}_u=  \\ \notag
&&\frac{({\bf u},{\bf u})^2}{2}
\left\{-f(R)u{\bf e}_u-4g(\alpha,R)({\bf u},{\bf u})^{-3/2}
({\bf u}^{\prime},{\bf u}^{\prime})^{1/2} 
[u'{\bf e}_u+u\phi'{\bf e}_{\phi}]\right\}
\end{eqnarray}
This yields the following two equations for the tangential and  
radial components
\begin{equation}
\label{4.8}
u\phi''+2u'\phi'=-2g(\alpha,R)u^2
({\bf u}^{\prime},{\bf u}^{\prime})^{1/2}\phi',
\end{equation}
\begin{equation}
\label{4.9}
u''-u(\phi')^2-\frac{({\bf u}',{\bf u}')}{u}=-\frac{f(R)u^5}{2}-2g(\alpha,R)u
({\bf u}^{\prime},{\bf u}^{\prime})^{1/2}u',
\end{equation}
where
$$
({\bf u}',{\bf u}')=(u')^2+u^2(\phi')^2.
$$
Multiplying (\ref{4.8}) by $u$, dividing by $u^2\phi'$ and integrating 
we obtain
\begin{equation}
\label{4.10}
L=u^2\phi'=h_1\exp\left(-2\int g(\alpha,u)
u({\bf u}^{\prime},{\bf u}^{\prime})^{1/2}\,ds\right).
\end{equation}
Hence 
\begin{equation}
\label{4.11}
\frac{d}{ds}=\frac{L}{u^2}\frac{d}{d\phi}
\end{equation}

We note that $L$ defined in (\ref{4.10}) is equal to the angular momentum
$J$ up to a constant. In fact using (\ref{4.2}) we have
$$
L=u^2\phi'=u^2R\frac{d\phi}{dt}=\frac{R^2{\dot\theta}}{2}
$$

Using(\ref{4.11}) to change the variable from $s$ to $\phi$ in (\ref{4.9})
we obtain after a long algebra the following orbit equation,
\begin{equation}
\label{4.12}
\frac{1}{u}\frac{d^2u}{d\phi^2}-\frac{2}{u^2}\left(\frac{du}{d\phi}\right)^2
-\frac{u^2({\bf u}^{\prime},{\bf u}^{\prime})^{1/2}}{L^2}=
1-\frac{f(u)u^8}{2L^2}.
\end{equation}
When $f(R)$ is given by (\ref{3.6}) this becomes
\begin{equation}
\label{4.13}
\frac{1}{u}\frac{d^2u}{d\phi^2}-\frac{2}{u^2}\left(\frac{du}{d\phi}\right)^2
+\frac{u^2[\mu-2({\bf u}^{\prime},{\bf u}^{\prime})^{1/2}]}{2L^2}=1
\end{equation}
Using (\ref{3.15}) this equation becomes
\begin{equation}
\label{4.14}
\frac{1}{u}\frac{d^2u}{d\phi^2}-\frac{2}{u^2}\left(\frac{du}{d\phi}\right)^2
-\frac{Eu^4}{2L^2}=1
\end{equation}
When there are no drag forces $E$, $L$ are constants and eq. (\ref{4.14})
may be referred to as the "Energy Angular Momentum" orbit equation in 
Levi-Civita coordinates. In absence of dissipative forces this equation 
characterize the orbit by its generic invariants. 

\setcounter{equation}{0}
\section{Relative motion of Satellites}

In this section we derive the equations for the relative motion of two
satellites in orbit in a central force field.

\subsection{Linearized Equations in Physical Space}

If the positions of the two satellites are denoted by
${\bf R_1},{\bf R}_2$ then their respective equations of motion 
in a (conservative) central force field are
\begin{equation}
\label{5.1}
{\ddot{\bf R}}_1= -\nabla V({\bf R}_1),\;\;\; 
{\ddot{\bf R}}_2= -\nabla V({\bf R}_2)
\end{equation}
where the dots represent differentiation with respect to time. The relative 
position of the second satellite with respect to the first 
is ${\bf w}={\bf R}_2 -{\bf R}_1$ . This leads to the equation 
\begin{equation}
\label{5.2}
\ddot{R_1}+\ddot{\bf w}= -\nabla V({\bf R}_2).
\end{equation}
Using (\ref{5.1}) this can be rewritten as
\begin{equation}
\label{5.3}
\ddot{\bf w}= \nabla V({\bf R}_1)-\nabla V({\bf R}_2).
\end{equation}
Assuming that $|{\bf w}| \ll R_1$ we can approximate 
$$
\nabla V({\bf R}_1)-\nabla V({\bf R}_2) = 
\nabla \left[V({\bf R}_1)- V({\bf R}_1+{\bf w})\right]
$$
by a first order Taylor polynomial in ${\bf w}$. 
This leads to the following linear relative motion of the second satellite  
with respect to first in the inertial coordinate system attached to the 
central body center,
\begin{equation}
\label{5.4}
\ddot{\bf w}=-\nabla(\nabla V(R_1)\cdot {\bf w}) 
\end{equation}
In particular if the motion is around a spherical body where $V$ is given
by (\ref{3.6}) we have
\begin{equation}
\label{5.4a}
\ddot{\bf w}= -\frac{\mu}{R_1^3}{\bf w}+
\frac{3\mu({\bf R_1}\cdot {\bf w})}{R_1^5}{\bf R_1}={\bf F}.
\end{equation}

In a coordinate system rotating with the first satellite the relative-motion
equation
(\ref{5.4}) becomes [21]
\begin{equation}
\label{5.5}
\ddot{{\bf w}} +2\bfOmega \times \dot{\bf w}+\bfOmega \times 
(\bfOmega \times {\bf w}) +\dot{\bfOmega} \times {\bf w} = {\bf F}.
\end{equation}
where $\bfOmega$ is the orbital angular velocity of the first satellite.
The reduction of this formula to a system of ordinary differential equations
for the motion of two satellites around an oblate body was carried in [18].

We now consider this equation in the special case where the two satellites are 
in the same $x-y$ plane. In this case
\begin{equation}
\label{5.6}
\bfOmega=(0,0,{\dot \theta}),\,\,\,\, {\bf w} =(w_1,w_2,0).
\end{equation} 
We have
$$
\bfOmega \times {\dot{\bf w}}={\dot{\theta}}(-{\dot w}_2,{\dot w}_1,0)^T,\,\,\,
\bfOmega \times (\bfOmega \times {\bf w})=-{\dot \theta}^2(w_1,w_2,0)^T,\,\,\,
{\dot{\bfOmega}}\times {\bf w}={\ddot \theta} (-w_2,w_1,0)^T
$$
(In the following we suppress the third component of the vectors).

\subsection{Relative Equation of Motion in Levi-Civita Coordinates}

We now introduce Levi-Civita transformation
\begin{equation}
\label{5.7}
w_1=v_1^2-v_2^2,\,\,\, w_2=2v_1v_2,\,\,\, r^2=w_1^2+w_2^2=
({\bf v},{\bf v})^2,\,\,\, \frac{d}{ds}={r}\frac{d}{dt}.
\end{equation} 

Due to the appearance of the vector $(-w_2,w_1)$ in the equation of motion
(\ref{5.5}) we introduce 
\begin{eqnarray}
\label{5.8}
{\bar L}({\bf u})=\left(\begin{array}{cc}
-u_2 & -u_1 \\ \notag
u_1 &  -u_2 
\end{array}\right).
\end{eqnarray}
We then have
\begin{equation}
\label{5.9}
{\bar L}({\bf u}){\bf u}=(-u_2,u_1)^T.
\end{equation}
The matrix ${\bar L}({\bf u})$ has the following properties
\begin{equation}
\label{5.10}
{\bar L}({\bf u})^T{\bar L}({\bf u})=({\bf u},{\bf u})I,\,\,\,
{L}({\bf u})^T{\bar L}({\bf u})=({\bf u},{\bf u})\Gamma,\,\,\,
{L}^{-1}({\bf u})^T{\bar L}({\bf u})=\frac{1}{({\bf u},{\bf u})}
{L}({\bf u})^T{\bar L}({\bf u})=\Gamma,
\end{equation}
where 
\begin{eqnarray}
\label{5.11}
\Gamma=\left(\begin{array}{cc}
0 & -1 \\ \notag
1 &  0 
\end{array}\right).
\end{eqnarray}
Observe also that
\begin{eqnarray}
\label{5.12}
\left(\begin{array}{c}
-w_2' \\ \notag
w_1' 
\end{array}\right)=
2{\bar L}(v){\bf v}^{\prime}.
\end{eqnarray}
After some algebra similar to the the one in Sec $4$ we obtain the 
following representation of eq. (\ref{5.5})
\begin{equation}
\label{5.13}
{\bf v}^{\prime\prime}-\left[\frac{({\bf v}',{\bf v}')}{({\bf v},{\bf v})}+
\frac{r^2{\dot \theta}^2}{2}\right]{\bf v}+2r{\dot \theta}\Gamma{\bf v}'+
\frac{r^2{\ddot \theta}}{2}\Gamma{\bf v}=\frac{r^2L^T(v){\bf F}}{2}.
\end{equation}
Using
$$
{\dot{\theta}}=\frac{\theta'}{r},\,\,\,
{\ddot \theta}=\frac{1}{r^2}\left[\theta''-\frac{2}{r}
({\bf v},{\bf v}')\theta'\right].
$$
eq (\ref{5.13}) becomes
\begin{equation}
\label{5.14}
{\bf v}^{\prime\prime}-\left\{\frac{({\bf v}',{\bf v}')}{({\bf v},{\bf v})}+
\frac{(\theta')^2}{2}-\frac{1}{2}\left[\theta''-\frac{2({\bf v},{\bf v}')}
{({\bf v},{\bf v})}\theta'\right]\Gamma\right\}{\bf v}+
2(\theta)'\Gamma{\bf v}'=r^2\frac{L^T(v){\bf F}}{2}.
\end{equation}

In particular when $\bf F$ is given by (\ref{5.4a}) the right hand side
of this equation becomes
$$
r^2\frac{L^T({\bf v}){\bf F}}{2}= -\frac{\mu r^3}{2R_1^3}{\bf v}+
\frac{3\mu r^2}{2R_1^5}({\bf R_1}\cdot {\bf w})L^T({\bf v}){\bf R_1}
$$

\setcounter{equation}{0}
\section{Numerical Verification}

\subsection{Motion of a Satellite in an Exponential Atmosphere}

A common model for the earth atmosphere density $\rho$ with height is
\begin{equation}
\label{6.1}
\rho= C_1\exp\left(\frac{R_0-R}{H}\right)
\end{equation}
where $C_1,\; R_0,\; H$ are constants. For this atmospheric model
\begin{equation}
\label{6.2}
g(\alpha,R)=\alpha\exp\left(\frac{R_0-R}{H}\right)=
\alpha\exp\left(\frac{u_0^2-u^2}{H}\right)
\end{equation}
where the constant $C_1$ was lumped with the drag coefficient $\alpha$.
Eq. (\ref{3.7a}) for the energy becomes
\begin{equation}
\label{6.3}
\frac{dE}{ds} = -\frac{8\alpha}{u}({\bf u}'\cdot {\bf u}')^{3/2}
\exp\left(\frac{u_0^2-u^2}{H}\right).
\end{equation}
Similarly for $L$ we have 
\begin{equation}
\label{6.4}
\frac{1}{L}\frac{dL}{ds} = -2\alpha u({\bf u}'\cdot {\bf u}')^{1/2}
\exp\left(\frac{u_0^2-u^2}{H}\right).
\end{equation}
Using (\ref{4.11}) to change variables from $s$ to $\phi$ in (\ref{6.3}), 
(\ref{6.4}) yields,
\begin{equation}
\label{6.5}
\frac{dE}{d\phi}=
-\frac{8\alpha L^2}{u^5}
\left(\frac{d{\bf u}}{d\phi}\, ,\, \frac{d{\bf u}}{d\phi}\right)^{3/2}
\exp\left(\frac{u_0^2-u^2}{H}\right)
\end{equation}
\begin{equation}
\label{6.6}
\frac{dL}{d\phi}=
-2\alpha uL
\left(\frac{d{\bf u}}{d\phi}\, ,\, \frac{d{\bf u}}{d\phi}\right)^{1/2}
\exp\left(\frac{u_0^2-u^2}{H}\right)
\end{equation}
The system of equations (\ref{4.14}),(\ref{6.5}),(\ref{6.6}) comprise of 
four equation in four unknowns and can be solved by numerical methods.
Fig $1$ is a plot of the solution of this system with $R_0=7000km$,\,
${\dot \theta_0}=\sqrt{\frac{\mu}{R_0^3}}$ (initial circular orbit),
$\alpha=3.10^{-10}$, $H=88.667$ and step error of $10^{-12}$. On the 
same figure we plotted also the numerical solution of the system 
(\ref{2.8})-(\ref{2.9}) with the same parameters. We note that these
curve are almost inditinguishable. Fig $2$ displays the difference 
between these curves which over ten periods remains less than $0.75m$
This difference is most probably due to the cumulative error in the 
numerical integration.

To verify numerically the formula for the relative motion of satellites
which was derived in the previous section we considered two satellites
in circular orbit whose position at time $t=0$ (in polar coordinates) is 
$(R_1,0)$ and $(R_2,0)$ with $R_1=7000 km$ and $R_2=6999 km$. The angular 
velocities of these satellites respectively are
$$
\omega_i={\dot \theta}_i=\sqrt{\frac{\mu}{R_i^3}}  ,\,\,\,\, i=1,2
$$
Hence their distance $d$ at time $t$ satisfies
$$
d^2 = R_1^2+R_2^2-2R_1R_2\cos(\omega_1-\omega_2)t
$$ 
Fig $3$ is a plot of the difference between this analytical expression 
for the distance and the numerical value obtained from the linearized 
formula (\ref{5.13}) as a function of time.

\section{Conclusion}

In the first part of the paper we developed a new representation for
the orbit equation of a satellite in terms of it natural invariants.
In the second part a formula was derived for the relative motion of two 
satellites moving in the same plane. This formula can be generalized to 
the case where drag effects have to be taken into account and the orbits
of the satellites are not in the same plane (using KS-formalism).

\newpage
\begin{figure}[ht]
\centerline{\includegraphics[height=100mm,width=120mm,clip,keepaspectratio]{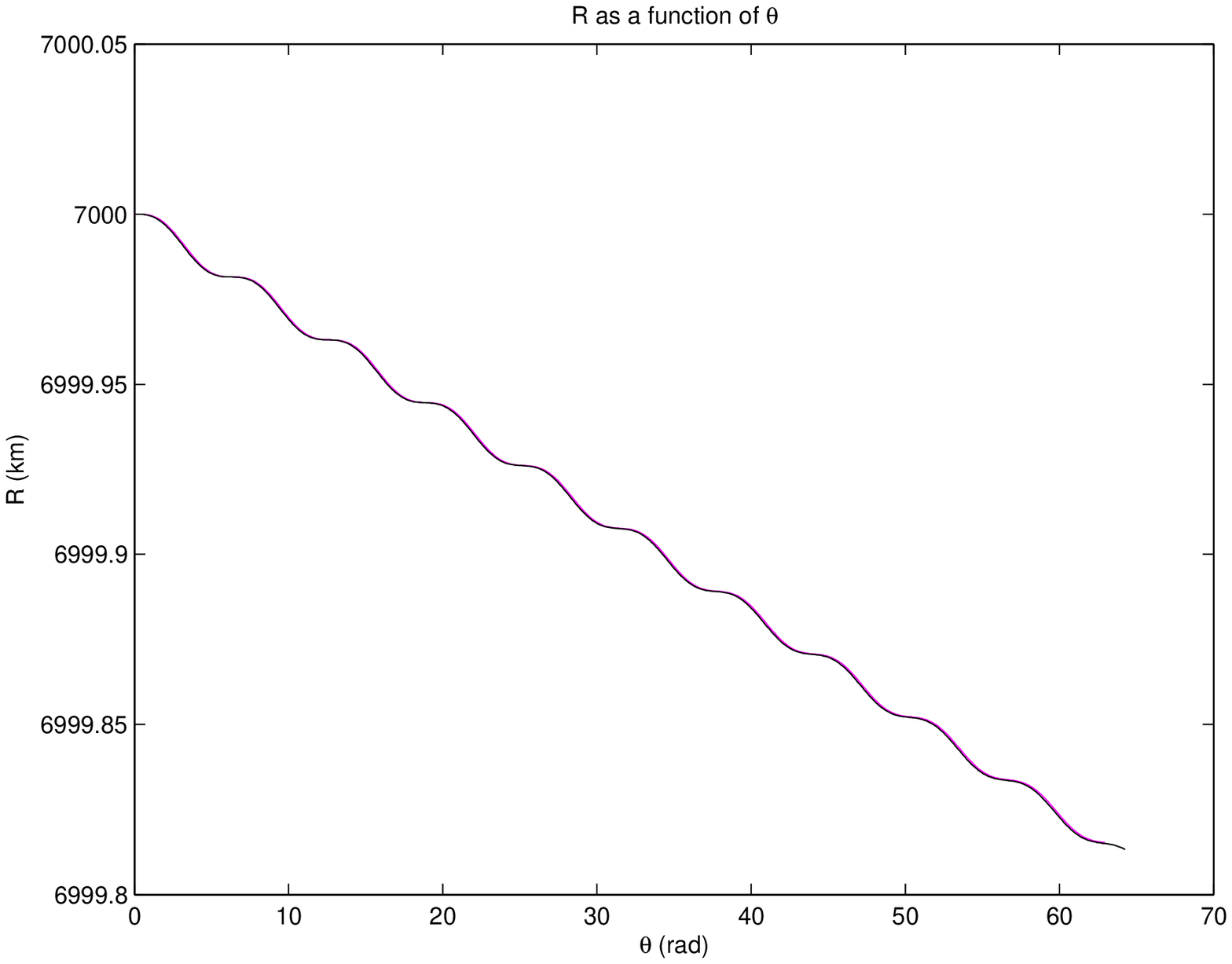}}
\label{Figure 1}
\caption{:\,Illustrative trajectory for a satellite orbit using eq. 
(\ref{4.14}) (red line) which is indistinguishable from the one obtained from
(\ref{2.8})-(\ref{2.9}).}
\end{figure}

\newpage
\begin{figure}[ht]
\centerline{\includegraphics[height=100mm,width=120mm,clip,keepaspectratio]{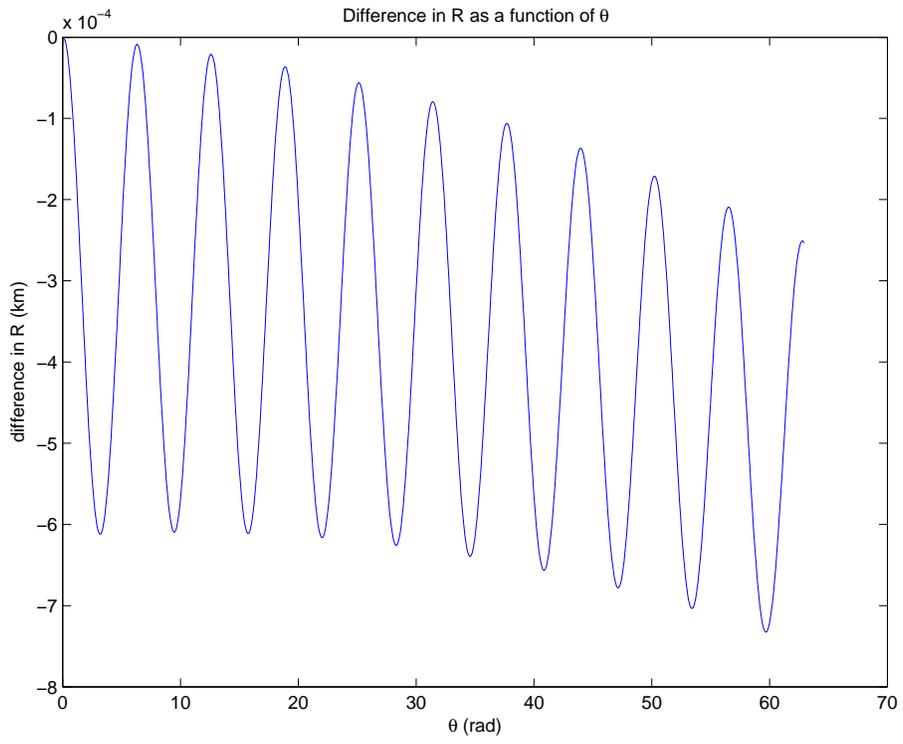}}
\label{Figure 2}
\caption{:\,Diffrence between the trajectories in Fig $1$}
\end{figure}

\newpage
\begin{figure}[ht]
\centerline{\includegraphics[height=100mm,width=120mm,clip,keepaspectratio]{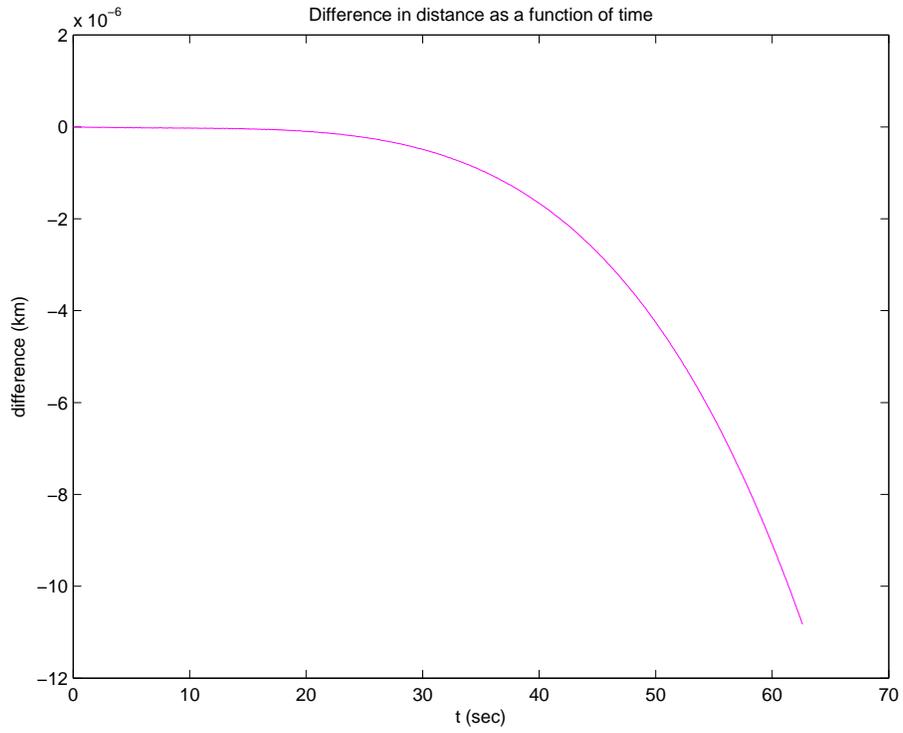}}
\label{Figure 3}
\caption{:\,Diffrence between the analytic and numerical value of the
distance between two satellites}
\end{figure}

\end{document}